\def\year{2019}\relax

\documentclass[letterpaper]{article} 
\usepackage{aaai19}  
\usepackage{times}
\usepackage{helvet}
\usepackage{courier}
\usepackage{balance}  
\usepackage{graphicx} 
\usepackage{url}      
\usepackage{subfig} 
\usepackage{caption} 
\usepackage{enumitem}
\usepackage{amsthm,amssymb, amsmath}

\newlist{hypotheses}{enumerate}{1}
\setlist[hypotheses, 1]{label = H.\arabic*:}

\nocopyright
\def\plaintitle{ Let's Keep It Safe: Designing User Interfaces that Allow Everyone to Contribute to AI Safety}
\def\plainauthor{Travis Mandel, Jahnu Best, Randall H. Tanaka, Hiram Temple, Chansen Haili, Kayla Schlectinger, Roy Szeto}

\usepackage[pdflang={en-US},pdftex]{hyperref}
\newtheorem{theorem}{Theorem}
\newtheorem{assumption}{Assumption}

\DeclareMathAlphabet{\mathcal}{OMS}{cmsy}{m}{n}

\begin{document}
%

\DeclareRobustCommand{\okina}{%
  \raisebox{\dimexpr\fontcharht\font`A-\height}{%
    \scalebox{0.8}{`}%
  }%
}

\title{\plaintitle}

\author{
Travis Mandel$^1$, Jahnu Best, Randall H. Tanaka$^1$, Hiram Temple$^1$, Chansen Haili$^1$, Kayla Schlectinger$^1$$^2$,\\ \textbf{ \Large Roy Szeto}$^3$\\
$^1$Computer Science Department, University of Hawai{\okina}i at Hilo, Hilo, HI\\
$^2$ Computer Science and Engineering, University of Minnesota, Minneapolis, MN\\
$^3$Center for Game Science, University of Washington, Seattle, WA\\
\{tmandel, jahnub, dh404, htemple, haili808, kayla82\}@hawaii.edu, roylszeto@gmail.com
}

\maketitle
\begin{abstract}
When AI systems are granted the agency to take impactful actions in the real world, there is an inherent risk that these systems  behave in ways that are harmful. Typically, humans specify constraints on the AI system to prevent harmful behavior; however, very little work has studied how best to facilitate this difficult constraint specification process. In this paper, we study how to design user interfaces that make this process more effective and accessible, allowing people with a diversity of backgrounds and levels of expertise to contribute to this task.   We first present a task design in which workers evaluate the safety of individual state-action pairs, and propose several variants of this task with improved task design and filtering mechanisms.  Although this first design is easy to understand, it scales poorly to large state spaces. Therefore, we develop a new user interface that allows workers to write constraint rules without any programming.  Despite its simplicity, we show that our rule construction interface retains full expressiveness.   We present experiments utilizing crowdworkers to help address an important real-world AI safety problem in the domain of education.  Our results indicate that our novel worker filtering and explanation methods outperform baseline approaches, and our rule-based interface allows workers to be much more efficient while improving data quality.

\end{abstract}


\section{Introduction}

As AI systems become more ubiquitous, and are given greater agency to impact human lives, we must ensure that these systems behave in a \textit{safe} manner.  Although when one hears the term ``AI safety'' it brings to mind robot uprisings and autonomous cars running down pedestrians, this term is broad enough to encompass types of harm that are not physical in nature. One recent example of this is online movie recommendations~\cite{Balakrishnan2018nflix}, where we do not want an AI system to recommend a movie with extreme violence to a young child (even if they are likely to watch it). Another related example is education, where we want to empower AI systems to select among a wide array of content in order to personalize the educational experience for each student, but we do not want to allow the AI to teach incorrect or misleading information to impressionable young students.

A standard machine learning approach to this problem is simply to learn through trial-and-error which actions are harmful. However,  actually trying a potentially unsafe action in the real system is extremely undesirable in most settings. Therefore, researchers have studied AI systems that learn to predict when actions are unsafe before they executed, either based on results from simulation~\cite{dosovitskiy2017carla} or from explicit feedback~\cite{awad2018moral,balakrishan2018,Balakrishnan2018nflix}. Unfortunately, this is difficult in many real-world systems (especially those without a high quality simulator) as if the agent makes a mistaken prediction it may execute unsafe actions. Therefore,  AI systems often rely on experts in both machine learning and the specific domain to hand-craft rules, called \textit{constraints}, that specify what behaviors are unsafe.  In this setting, AI researchers tend to focus on studying how to create machine learning systems that obey these prespecified constraints~\cite{dalal2018safe,chow2018lyapunov,tessler2018reward}.      

Although this research is valuable from an AI perspective, often the weakest link are the human factors.  Unfortunately, designing these constraint rules is time consuming and error prone, especially when the space of all states and actions is large and complex.    Additionally, the group of experts who write these constraints is limited and they may not share the same values or perspectives as the general populace.    The valuable time and effort expended to generate these constraints is also in competition with other important tasks related to improving the AI system (such as improving the underlying algorithm).   Although the core AI problem of integrating constraints is very well studied, there is unfortunately very little work exploring how to most effectively design the interaction process by which humans specify these constraints.

In this paper, we present the first study of how to develop user interfaces that facilitate effective constraint specification for AI systems.  Our goal is to make interfaces that are easy to use, allow a variety of individuals with diverse backgrounds to contribute, ease the burden on the limited group of experts, and ensure high-quality output. Therefore, we take a crowdsourcing approach to this work, evaluating our task designs on Amazon Mechanical Turk. Although we do not expect crowdworkers to have an in-depth understanding of the specific domain, for many AI problems, one does not need to be an expert to recognize when an AI system is about to execute an unsafe action.  For example, one does not need to be an expert event planner or child psychologist to recognize that booking a band known for their explicit lyrics for a child's birthday party is in some sense unsafe.

  In our first interface design, we visualize a situation and a candidate action to workers, and ask them whether or not the action would be safe to take in that context. Initial performance is surprisingly poor, so we explore new variations of the task design that produce higher quality constraints by better filtering workers and promoting more careful work.    Although this ``case-by-case'' interface design is relatively straightforward, it seems inefficient when there are a large number of state-action pairs.  Therefore, we additionally develop an interface that allows workers to write rules that can constrain many state-action pairs at the same time.  Our rule-based interface does not require any programming, and gives users near-instant feedback about the impact of the rule.  Further, we develop a novel simplification that makes the rule construction task more accessible, while provably retaining full expressiveness.  We evaluate our scenarios on an important real world AI problem in education, using data from thousands of students and several educational experts.  We find that our new methods of filtering and promoting careful work are effective at increasing precision. We also find that our rule-based approach improves upon the case-by-case approach in terms of precision, while being much more cost-efficient.

\section{Related work}
\subsection{Crowdsourcing Decisions and Constraints}
Perhaps the most closely related to our paper is recent work by Awad et al.~\citeyear{awad2018moral} on using large amounts of non-expert workers to determine which decisions are most ethical for an AI system.  This work asks the user to choose between multiple unsafe actions (i.e. killing passengers or killing pedestrians) in artificially constructed scenarios. However, Awad et al. did not explore how best to design the user interface for this task, and makes the assumption that users always select their true preference. In our work, we focus on determining the safety of actions in real-world scenarios, and even though the correct answer is much more clear cut in our domain, we observe that many workers struggle to understand this difficult task. This motivates our exploration of how to design user interfaces to increase data quality in this safety-critical setting. 


Work by Zhuo~\citeyear{zhuo2015crowdsourced} (and Gao et al.~\citeyear{gao2015acquiring}) looks at a related problem, in which they ask human workers to help specify an action model (including constraints) for various simulated AI planning tasks.  
However, instead of having workers build their own constraints, they simply had users answer whether or not a given pre-generated constraint was correct.  Also, note that constraints are not a safety issue in planning, unlike in our case where the AI plans to take these actions in the real world. As such, this work did not investigate user interface designs to improve worker performance.

Other work has focused on using human workers to add new actions to an AI system \cite{williams2016axis,mandel2017where}.
Indeed, related past work has looked at using crowdworkers to write hints which could be added to an educational game, potentially serving as actions for a downstream AI system \cite{chen2016crowdsourcing}. However, these crowdworker-written hints were not found to be high-quality enough to deploy in-game without needing extensive expert filtering and revision.  Therefore, in this paper we assume the task of writing hints (developing actions) is handled by education experts. But these experts have very limited time and resources to write new hints, and as such we focus on how to design accessible tasks that maximize the efficacy of these hints by determining in what situations it is safe to try them.


\subsection{Data Quality}
Vast amounts of past work has looked at different procedures to increase the accuracy of data coming from crowdworking platforms such as Mechanical Turk~\cite{soylent,ambati2012,heer2010,mitra2015comparing}.  However, in this task our goal is not simply to have high accuracy, but rather to have extremely high precision and non-negligible recall, which is a very different objective and requires changes in the task design to achieve.  Further, these works have looked at improving accuracy on other types of tasks such as image labeling or sentiment analysis, and therefore cannot leverage or examine any specific properties of constraint-design tasks.\footnote{For example, the fact that we need a large amount of annotations on a single action in different locations, or that we have pre-built gold knowledge based on default system actions.}  As we explore throughout the rest of the paper, real-world AI constraint design tasks are particularly difficult to make accessible for a variety of reasons.

Although there is a large body of work exploring the quality and impact of gold questions for the purpose of filtering crowdworkers~\cite{dai2011artificial,bragg2016optimal}, most of it requires experts to expend substantial additional effort annotating a sizable dataset of positive and negative examples. Some work has studied programmatically generating gold questions \cite{oleson2011programmatic}, however it still requires experts to spend significant time identifying common mistakes and construct mutation operators and associated error descriptions. In our setting we get certain gold labels without needing to expend this effort, and we investigate whether filtering workers based on these alone is sufficient to ensure high precision.




\section{Problem Setup}
\label{sec:probsetup}
We deal with problems where an AI system, known as an \textit{agent}, must take actions in an unknown real-world environment.    Specifically, we assume the agent has a set of actions $\mathcal{A}$ and a (possibly large) set of states $\mathcal{S}$.\footnote{This is common formalism in AI. States can be though of as ``situations'' or ``contexts'', and actions can be though of as ``interventions'' or things an agent can do to affect the world around it.} The environment produces a state $s \in \mathcal{S}$, the agent chooses an action $a \in \mathcal{A}$, and the environment transitions to a new state $s'' \in \mathcal{S}$ . We assume there exists a function $C(s,a)$, which is unknown to the agent, that returns true if and only if the action $a$ is safe to take at state $s$.  In this paper, our goal is to use humans (e.g. crowdworkers) to find a Boolean function $\hat{C}(s,a)$ that approximates $C(s,a)$ as well as possible, so that we can develop ``safe'' agents that only take actions where $\hat{C}(s,a)=true$.   Other than this,  we are completely agnostic to the details of the AI algorithm the agent uses to select actions (this could include reinforcement learning approaches, recommendation algorithms, etc.). 

In terms of evaluating $\hat{C}(s,a)$, our primary aim is to maximize \textbf{precision}; that is, the proportion of cases where $\hat{C}(s,a) = true$ that  $C(s,a) = true$. This is because, since we intend to feed  $\hat{C}(s,a)$ to an autonomous agent, if $\hat{C}(s,a) = true$ and $C(s,a) = false$, the agent may take an unsafe action. Of course, one can get high precision just by letting $\hat{C}(s,a)=false$ almost everywhere, so the number of cases where $\hat{C}(s,a) = true$ is a secondary consideration.  
We make the following simplifying assumption:
\begin{assumption}
For every action $a'$,  we can produce at least  one state  $s'$ such that $C(s',a')=true$; in other words, a state where $a'$ is known to be safe.
\label{ass:posgold}
\end{assumption}
This assumption is not very strong, since usually actions are constructed with a specific situation in mind. For example, a robot may have a manipulator designed to pick up a certain type of block, or a telemarketing agent may have a standard script giving the normal situation a line should be uttered in.  In situations where this is not the case, we imagine that it should be relatively easy for an expert to identify some state where the action is safe to try.\footnote{If it is hard to identify any states where the action is safe, one wonders why the agent was provided this action in the first place!}

\section{Case-by-case Design}
\begin{figure}
    \centering
    \includegraphics[width=\columnwidth]{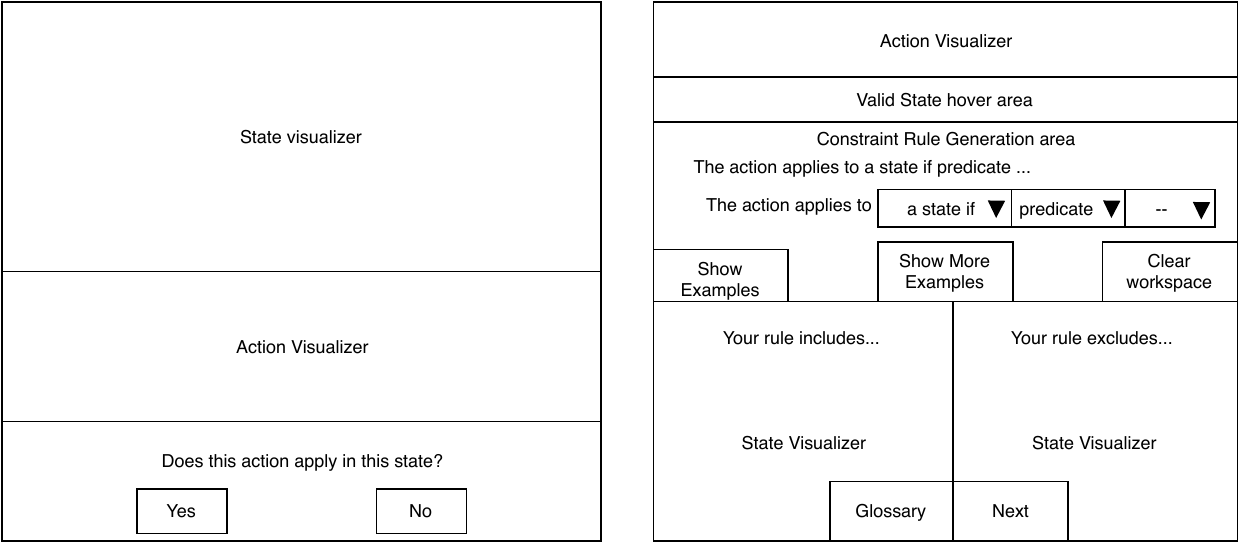}
    \caption{A comparison of the general-purpose design of our case-by-case interface (left) and our rule-based interface (right).}
    \label{fig:interface_gen}
\end{figure}
\label{sec:casebycase}
One way to dramatically simplify the task of constructing the constraint function $C(s,a)$ is to have human workers view individual $(s,a)$ pairs and give a Boolean response indicating whether action $a$ is safe  to try in state $s$. 



Figure 1 shows the basic task design. We visualize the state to the user, show the action, and then allow the user to respond (yes/no) as to whether the action applies.

This task has the advantage of resembling a standard binary labeling task, which is familiar to many people (especially crowdworkers). However, it differs in a few key ways.  First,  in our case, we care much more about precision, compared to a typical binary labeling task where precision is roughly as important as recall (reducing false negatives). 
Although controlling the tradeoff between precision and recall is trivial in machine learning systems, it is unclear how this can be done with human workers. 
Second, this task seems quite complex compared to other crowdsourced tasks. Figure~\ref{fig:interface_gen} is somewhat deceptive in this respect, as the worker must consider numerous elements of the state and action, and think carefully about how they relate to each other.  
The final difference relates to gold questions, which are questions with a known response that can be interspersed in a task to filter out poorly-performing workers.  
To achieve quality responses, experts typically generate a large dataset of positive and negative gold questions, which  is a significant burden.  However, due to Assumption \ref{ass:posgold}, for every action $a''$ we have a single state $s''$ where that action is known to be safe.  Since the correct answer when presented with the pair $(s'', a'')$ is ``yes'', this is a gold question which requires no additional expert effort to generate.

Therefore, as a baseline way of filtering workers, we include a single positive gold question per task (randomly generated as per Assumption \ref{ass:posgold}), and  if workers answer no to any of these gold questions, their responses are excluded.  To train workers, we include a short tutorial  before the task, which simply explains to the worker why each response they submit is correct or incorrect.

\subsection{Filtering and Training}
 \label{sec:stage1cond}
To ensure high-quality work, effective worker training and filtering is key.  Therefore, we try several alternative methods to the previously-described \textbf{Baseline}: 

 \textbf{Tutorial Overload} The first time a worker performs the task, all but one question is a tutorial.  We generated the additional tutorial questions by going through randomly-selected state-action pairs and writing brief explanations for the yes and no answers.  If the original number of unknown questions in the first task is $N$, the expert will need to label $N-1$ gold questions.  In our experiment, of the 5  additional questions, 3 were yes answers and 2 were no answers.  

 \textbf{Gold Overload} The first time a worker performs the task, all but one question is a gold question.  The positive gold were generated using Assumption \ref{ass:posgold}, but the negative gold were generated by labeling $N-1$ gold questions as per Tutorial Overload.
Unfortunately, this means that the negative gold state-action pairs will always be the same and thus may be susceptible to memorization by workers.

 \textbf{Fake Gold} \label{sec:fakegold} Although in our application we found sufficient ``no'' answers fairly quickly, that is not a guarantee in general. In many cases, the actions are safe almost everywhere (e.g. a servomechanism on a robot, or the brake on a car), and have a relatively small number of situations where they are unsafe.  Finding these situations to generate more tutorial or gold questions places a tremendous burden on the expert.
 Furthermore, although the aforementioned methods are natural ways to test the impact of tutorials and gold on precision, they increase the cost per valuable worker answer substantially, as only one question in the initial task actually deals with a $(s',a')$ such that $C(s',a')$ is unknown. 
Therefore, to further reduce expert effort we introduce a ``fake'' negative gold action, i.e. a synthetic action that will clearly not apply to any states.  To make the action better test whether the workers understand the task,  we propose having the action refer to the task itself. 
For example, if the action space is text-based and we are running the task on Mechanical Turk, the action could read ``Keep up the good work! You only have [n] questions left before you complete this HIT!'' Or, if the action space is visual advertisements, one could show an image giving the worker positive reinforcement as a misdirect. 

Note that this approach is \textbf{not} equivalent to just adding one negative gold question.  The state (which consumes most of the case-by-case display, see Figure \ref{fig:interface_gen}) changes every time the worker gets a fake gold question.  Thus even if they were able memorize the correct response to every state-action pair they had previously seen, it would not directly help them when receiving the next fake gold question.


\textbf{H.1} We hypothesized that Tutorial and Gold Overload would help precision but generate very little work, while Fake Gold would retain a lot of the precision benefit but generate more useful work due to the higher ratio of unknown questions.  

\subsection{Promoting Careful Thinking}
 \label{sec:stage2cond}
Next, we explored a variety of approaches designed to further improve performance by encouraging workers to think carefully about relationship between states and actions:

 \textbf{Continuity}  Work by Lasecki et al. \citeyear{lasecki2014using} has studied workflow continuity, showing that presenting closely related tasks in succession can yield improved performance compared to a series of unrelated tasks (likely by reducing cognitive load).  In our case,  we keep the state the same throughout the task, allowing only the action to change. 
 To make detecting what had changed easier on the worker, we added special highlights to show that the action changes from task to task.  

\textbf{Skip} Past work \cite{stefano2015,chen2016crowdsourcing} has shown that promoting self-reflection is key to ensuring careful work. Therefore, we added a large ``Skip'' button between the ``Yes'' and ``No'' buttons, which is intended to cause workers to reflect on whether they are confident enough to submit a response, or would prefer to see a different question. To implement this button we pick another question of the same type (positive gold, fake gold, or non-gold), so workers still have to complete the same number of unknown and gold questions per task.  We also added a tool-tip explaining to the workers that there would be no penalty for pressing the skip button.  

\textbf{Two-sided Explanation} Past work~\cite{drapeau2016microtalk} has shown that requiring users to explain their answers can promote reflection and improve answer quality.  Therefore, we tried an approach where whenever the user selected an answer, a textbox appears in which they must type an explanation before the task can proceed.    Similar to past work~\cite{drapeau2016microtalk}, we apply a degree of filtering to the explanations: we do not allow the worker to proceed unless the explanation is 5th grade level (according to the Flesch-Kincaid scale) and at least 8 words. 
To further promote worker reflection, we added text and arrows that appear after an initial response is selected to make it clear workers could reconsider their response and choose the other option.


\textbf{One-sided Explanation}  A unique feature of our task is that, as mentioned in the problem setup section, we care primarily about reducing false positives.   Therefore, we developed an approach similar  to the two-sided explanation condition, except that when the user presses ``No'', no explanation is requested and they simply proceed to the next question. The hypothesis is that this approach would encourage low-effort workers to simply select ``No'', while encouraging high-effort workers to think more deeply about whether they have made the right selection. 

\textbf{H.2} Our hypothesis was that all the variants would improve precision. We felt our one-sided explanation condition would likely perform the best due to the combination of promoting reflection and filtering out low-effort workers.

\section{Rule-Based Method}
Our proposed case-by-case method scales poorly to larger state-action spaces, as workers must decide on every state-action pair individually.  The standard solution is to instead specify $C(s,a)$ by writing rules, which can forbid or allow a large set of state-action pairs. This is typically thought to be a task that requires extensive programming (in a language like LISP),  which greatly reduces the accessibility of the task.  Additionally, the process of defining a general rule (which encompasses arbitrary states and arbitrary actions) requires a comprehensive understanding of the state-action space, which is very challenging even for AI experts.  

Logically, there are two natural alternatives for writing more focused rules: Either a user is shown a single state and must write a rule defining what actions can safely apply there, or the user is shown a single action and must write a rule defining where (i.e., at which states) it is safe to apply that action.  The choice between these is somewhat domain-dependent, but in cases where the action space is large, action-specific rules seem the clear choice.  For example, a worker may have trouble understanding all the different visual advertisements that form the action space, but given a single advertisement likely has intuition about what types of customers it is applicable to.   This is also natural for systems that expand their capabilities (i.e. increase the action space) while keeping the state space fixed, which is common in domains such as education \cite{williams2016axis,mandel2017where}.  In these cases, before the new action(s) can be safely taken by the AI system, the system needs an action-specific constraint.  Therefore, we have users write rules that specify $C(s,a')$ for some chosen $a' \in \mathcal{A}$.

However, despite this simplification, the worker has to reason about the entire state space $\mathcal{S}$, which can be challenging.   To mitigate this, our system gives the user instant feedback about what states their rule includes and excludes.  

The layout of the task is shown in Figure \ref{fig:interface_gen}.  We place the action at the top of the screen, and below that we show the user (on hover) the known valid state for that action (which exists due to Assumption  \ref{ass:posgold}).  Below that is the rule-writing area, in which workers can use dropdowns to create a rule (described further in the next section).  When the user wishes to test their rule, they press the ``Show Examples'' button, which processes their rule and populates the included and excluded state areas with an visualized example of an included (or excluded) state.  If states are too large to visualize in the small space, we recommend showing some sort of condensed version and allowing it to expand when the user hovers over it. The user can then press ``Show More Examples'' if they want to see more example states, ``Clear Workspace'' to clear their rule, or ``Next'' to move on to writing a rule for the next action (after some basic validation).  Additionally, at all times, we allow workers to press a ``Glossary'' button to view a pop-up glossay window explaining the meanings of the different terms used.

\textbf{Dropdown-Based Rule Creation}
\label{sec:dropdown}
In order to enable workers to create expressive rules without programming, we designed a dropdown-based rule creation system.  The initial text says ``The action applies to'' and then the user is presented with a dropdown with the options ``all states'', ``no states'', or ``a state if''.  Selecting the first two terminates the rule, but selecting the last option generates another dropdown for the user to continue writing their constraint.  

 Constraints are usually specified in first order logic; more specifically, a logical combination of domain-specific predicates and arguments \cite{osullivan2002interactive,mitrovic2007intelligent}. Therefore, a natural approach is to represent these terms of dropdowns, where the user chooses a dropdown for a predicate followed by one or more dropdowns for the argument(s).  However, to a worker without background in AI this ordering is unintuitive, as in many natural languages (such as English) the predicate comes \textit{after} the argument(s), not before, for instance ``if door six is open'' instead of ``if isOpen(doorSix).'' 
Therefore we flip the order, asking users to select valid arguments first, and then auto-populating the next dropdown with the valid predicates for those argument(s). Additionally, if an argument only is valid for one predicate, we condense them into one dropdown to reduce confusion. 


After testing, a user may wish to change their rule. They can do this by pressing ``Clear Workspace'', or by changing any existing dropdown, at which point the following dropdowns are removed since the allowed follow-up dropdowns may have changed.

\textbf{Logic and Parentheses}
When writing constraints, it is important that the language is sufficiently expressive to allow users to write the desired constraint.  Clearly, this requires logical connectives, e.g. ``The robot is pointing at the target AND there is not a human in the way of the laser.''
Unfortunately this immediately raises the question of operator ordering and parentheses.  Take the following example: \textbf{(}the road is wet \textbf{AND} the car has hydroplaning-resistant tires\textbf{)} \textbf{OR} \textbf{(}the road is snowy \textbf{AND} the car has studded tires\textbf{)}

There is no way to present this expression exactly through left-to-right evaluation (in this example, left-to-right order would require studded tires for a wet road!).  Unfortunately, introducing parentheses complicates the task enormously, as the number of ways to parenthesize an expression grows roughly exponentially (Catalan) with the expression length.  

We developed a method to reduce the complexity of adding parentheses, which works by allowing users an occasional binary choice about where they want to place the next expression. 
As soon as a user introduces the first logical (AND or OR) such as ``A~OR'', it immediately adds the first set of parenthesis around the expression: 
``(A~OR~\dots)''. 

Once the user has completed the statement inside some set of parentheses, we create two special dropdown elements we call \textbf{\mbox{choiceboxes}} to allow them to choose where to put the next logical.  Specifically, there is a choicebox just inside the innermost right parenthesis, and a choicebox just outside the innermost right parenthesis, e.g. 
\mbox{``\dots~OR~D~--~)~--''} where the -- elements represent the choiceboxes. 
Upon choosing  a logical from one of the two choiceboxes, the unselected choicebox disappears. 
Additionally, a left parenthesis is placed just before the leftmost atomic element (which is a literal if they chose the innermost  choicebox, otherwise a parenthesized expression) and a space to add the next literal is added to the right of the new logical followed by a right parenthesis.  

Although this way to specify parentheses seems potentially more accessible, it would also seem to be quite limiting in terms of which expressions can be produced.  Perhaps surprisingly, this is not the case.  We show in Theorem \ref{thm:expressive} that our method is sufficiently expressive to represent any Boolean function of the predicates. The full proof is in the appendix section, here we provide a proof sketch.
\begin{theorem}
Aside from the length limit, our method of rule construction is fully expressive, that is, it can represent any Boolean function over the set of base predicates.
\label{thm:expressive}
\end{theorem}
\textbf{Proof Sketch}
 We show how our interface can allow the creation of any statement in disjunctive normal form (DNF). The basic idea is to select the outer choicebox in most cases, except when one is adding the second literal of a DNF clause, in which case one should add an inner choicebox in order to place a left parenthesis to separate out the clause. Since ANDs are always added at the outer choicebox, there is always at most one right parenthesis belonging to the clause and so it is easy to ensure that the OR is fully outside the clause. Finally, since any logical relation can be represented as a DNF~\cite{davey1990introduction}, the claim holds. 

Note that, although the proof works by showing any DNF expression is possible, it is straightforward to alter the proof to show that any CNF expression is possible as well.  We feel our parentheses method balances giving user choices about how to most intuitively organize logical expressions,  while keeping the interface simple and retaining full expressiveness.

\textbf{Tutorial \& Filtering Design}
Even with a simplified interface, the task is relatively complex, so an effective tutorial is key.  Our tutorial design works through a special ``Get Help'' button, which had text above it indicating that workers should press it if they were confused.  Pressing the button gives workers context-sensitive feedback on creating their rule, e.g. checking the number of included and excluded states to make sure they are roughly correct for that action.  If all those checks have passed, it shows the worker an expert-generated example rule and an associated explanation, and asks them to reconstruct it. 

Since each question is quite time-consuming, we do not include explicit gold questions to maximize the use of worker time.  Instead, to ensure high-quality work we filter out workers whose rule does not include the original (known valid) state. 

\textbf{H.3} Our hypothesis was that the rule-based interface would generate more positive responses than case-by-case due to the more efficient method of constraint specification, while retaining roughly equivalent accuracy.

\section{Experiment Setup}
As a real-world testbed for our approaches, we examine the AI problem of improving hints in an educational game, Riddle Books.  Riddle Books, developed by the Center for Game Science (CGS), has been played by over 350,000 people online. It teaches 3rd-5th grade students how to conceptually understand math word problems by diagramming out the relationship between items in the problem.   
At any time, students can receive a hint by pressing a hint button. A video demo of how the game works can be found in the video figure.

\subsection{The Riddle Books AI System}
The initial set of hints built into the game did not seem to be very effective. Therefore, CGS researchers recruited educational experts to help improve the hints. CGS showed the experts specific states where students were stuck and asked them to write a hint for each one.  These researchers plan to feed this dataset of hints to an AI system,\footnote{Specifically, one based on reinforcement learning.} which can learn over time which hints from the dataset are best for students in each situation.

The \textbf{action space} for the system consists of text-based hints written by the experts.  
In this paper we used a dataset of roughly 100 hints, but we expect the experts to continue writing hints and adding them to the system over time.

The \textbf{state space} for the system is defined by the level number and the student model (aka diagram) at which the hint button was pressed. Certain small alterations to a model (e.g. flipping two elements) are considered to be in the same state.
There are approximately 540 total states.

Before launching this AI system into the wild, properly defined constraints are key. Clearly, it would not be safe to simply launch the system with full freedom over showing any hint in any student state, as that would mean the game would show hints that might actually lead young children in the wrong direction (for example,  telling them something false, or giving them terrible advice).  CGS researchers tried an initial experiment only showing the new hints in the exact states the experts wrote them for. Unfortunately results were poor; likely due to the system being overconstrained.  With properly defined constraints, the system would have much greater freedom to safely try a single expert-written hint in many different student states, likely improving student outcomes.

To visualize our states,  we implemented a domain specific visualizer that renders states in a nearly identical manner to the way they are displayed in game.  
We developed a server-side backend system which automatically selects an exemplar diagram taken automatically from data of thousands of students playing Riddle Books on popular educational websites. This exemplar is then sent back to the client for visualization. 
Action visualization was trivial: we just displayed the text of a hint drawn from the expert hint dataset.

For the rule-based method, we created a total of 8 domain-specific predicates. These allowed rules to specify simple properties of the student model, for instance  saying ``the larger value is a bracket'' in the student model.  The rule is crafted on the client interface and sent to our server, which recursively processes it and quickly evaluates which student situations in the database it includes and excludes.

\begin{figure}
    \centering
    \includegraphics[width=0.8\columnwidth]{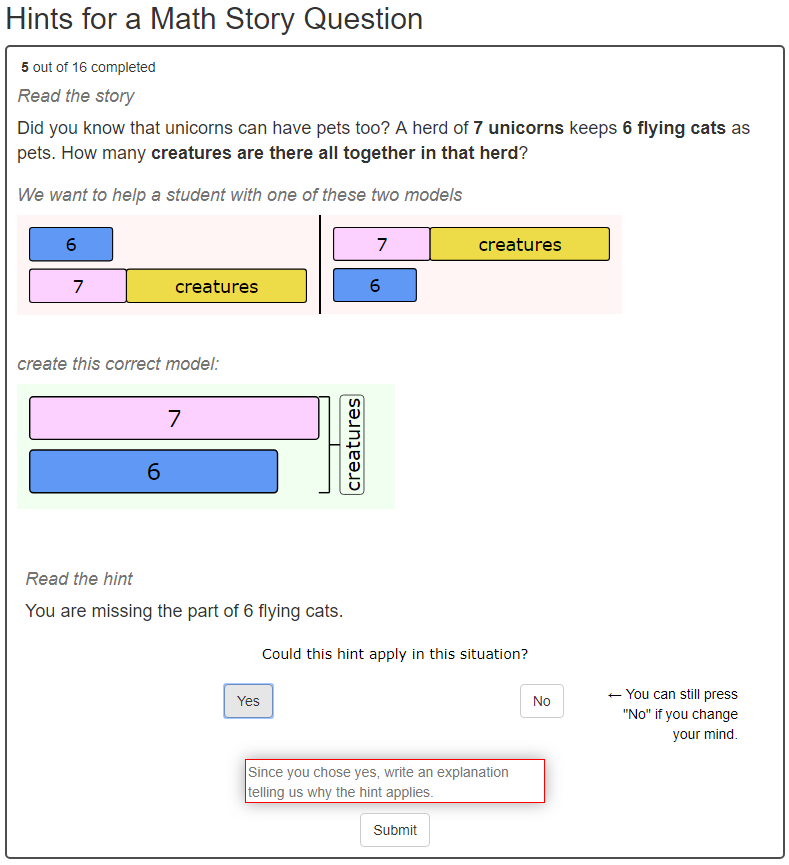}
    \caption{The Riddle Books case-by-case interface, with the prompt from the one-way explanation condition displayed. The correct answer is ``No'' in this case; the hint does not apply because the student already has a 6 block in their diagram to represent the flying cats.}
    \label{fig:rbcase}
\end{figure}
\begin{figure}
    \centering
    \includegraphics[width=\columnwidth]{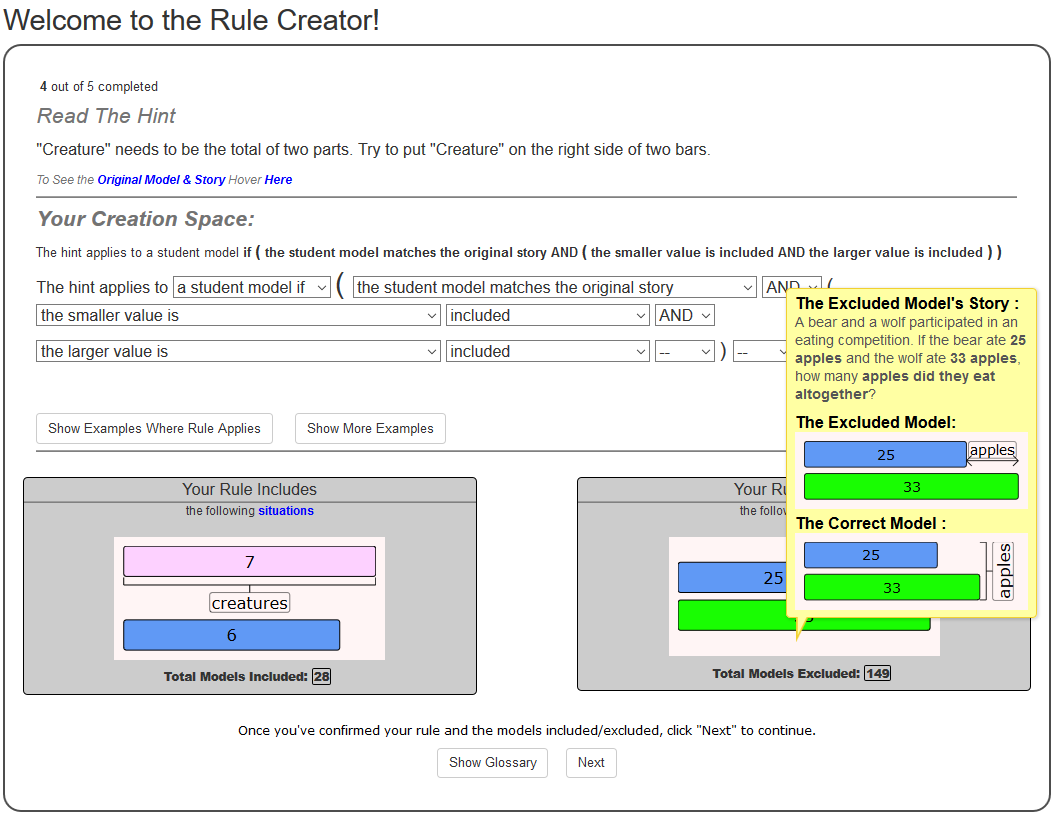}
    \caption{The Riddle Books rule creation interface, with a tooltip and choicebox displayed.}
    \label{fig:rbrule}
\end{figure}

Figure \ref{fig:rbcase} shows the Riddle Books case-by-case interface, while Figure  \ref{fig:rbrule} shows Riddle Books rule-based interface. A more dynamic presentation of our two interface designs can be found in the video figure. 

\subsection{Participants \& Procedures}
We launched our experiments on Amazon Mechanical Turk. 
In all experiments, workers were required to have a task approval rate of 98\% or higher along with at least one completed task, to ensure some small degree of filtering. 
Upon starting a HIT, the worker's Turk ID was associated with a randomly-assigned condition in our database, so that when a worker returns they are placed in the same condition.  Workers were required to agree to a standard consent form before starting the task, as required by our approved IRB protocol.  

In a small pilot study, we found one of the most complicated versions of the case-by-case task (one-sided explanation) to take roughly 5.5 minutes, thus we paid workers \$0.93 per HIT for the case-by-case experiments to ensure a reasonable hourly wage.
The first case-by-case HIT a worker completes is a 3-question tutorial followed by a 6-question task.  Subsequent case-by-case HITs are 7 non-tutorial questions.  To prevent our data from being overwhelmed by the results of a very small number of workers, we limit workers to a maximum of 5 case-by-case tasks.

Similarly, in a pilot study on Mechanical Turk we found the rule-based task to take roughly 21.5 minutes, thus we paid workers \$3.62 per HIT for the rule-based experiments to ensure a reasonable hourly wage. 
The first rule-based HIT a worker completes is a 2-question tutorial followed by a 3-question task.  Subsequent rule-based HITs are 4 non-tutorial questions.  We limit workers to a maximum of 3 rule-based tasks.

To evaluate precision, the first author of the paper judged the accuracy of the ``yes'' responses using an interface which blinded the conditions from which they were drawn. 
In all conditions the ultimate output consists of binary responses (whether an action applies to a specific state), but due to a low amount of samples in certain experiments we used the Fisher's exact test instead of the typical Chi-squared approximation. We use two-tailed tests unless otherwise noted.


\section{Results}

\begin{figure*}
\minipage{0.24\textwidth}
\includegraphics[width=\textwidth]{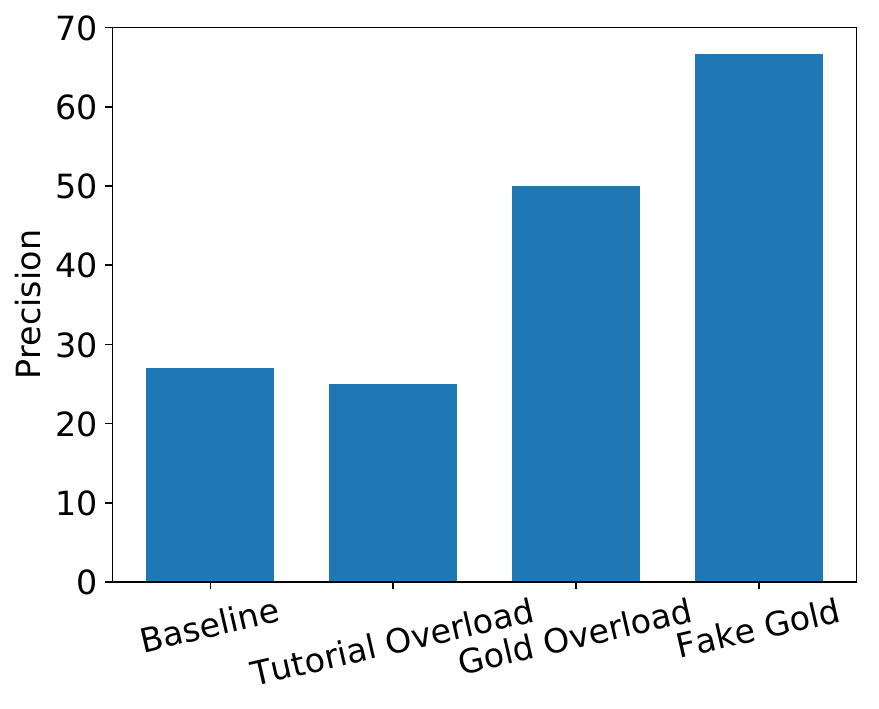}
\caption{Precision in our filtering and tutorial experiment.}
\label{fig:pre1}
\endminipage\hspace{0.1cm}
\minipage{0.24\textwidth}
\includegraphics[width=\textwidth]{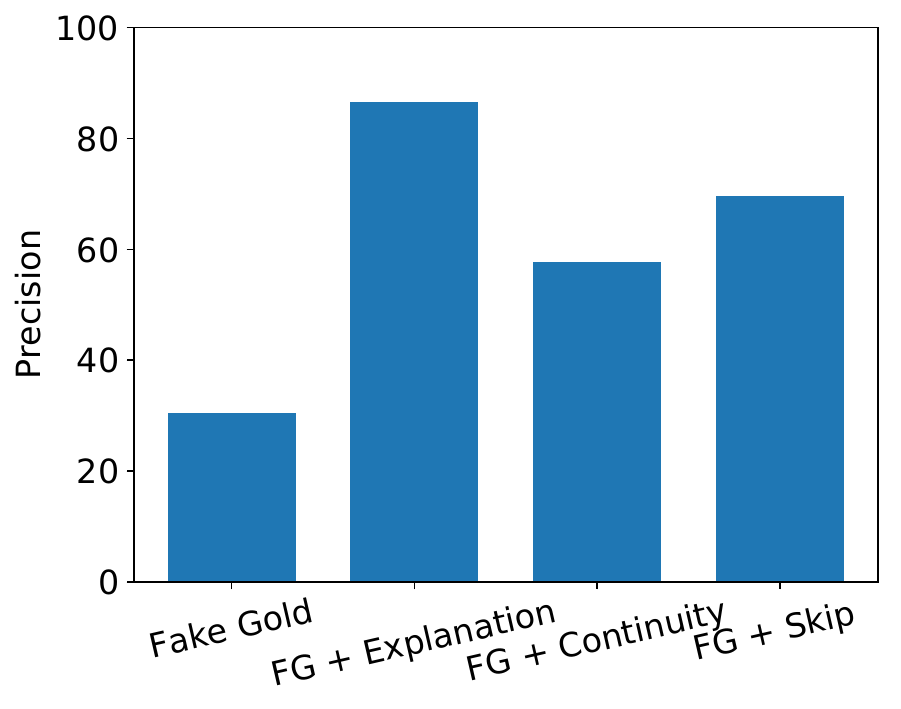}
\caption{Precision in our reflection design experiment.}
\label{fig:pre2}
\endminipage\hspace{0.1cm}
\minipage{0.24\textwidth}
\includegraphics[width=\textwidth]{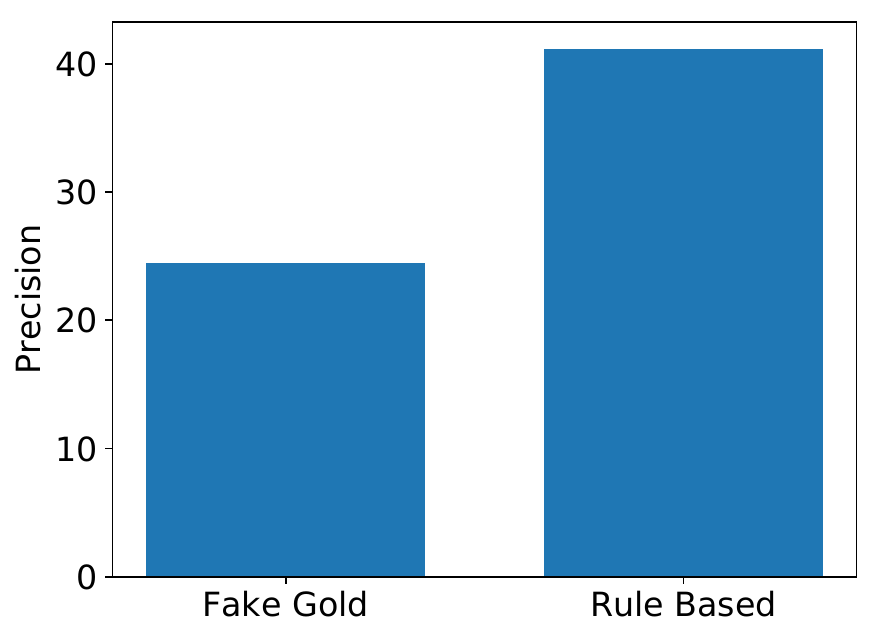}
\caption{Precision in our rule-based experiment.}
\label{fig:prerule}
\endminipage\hspace{0.1cm}
\minipage{0.24\textwidth}
\includegraphics[width=\textwidth]{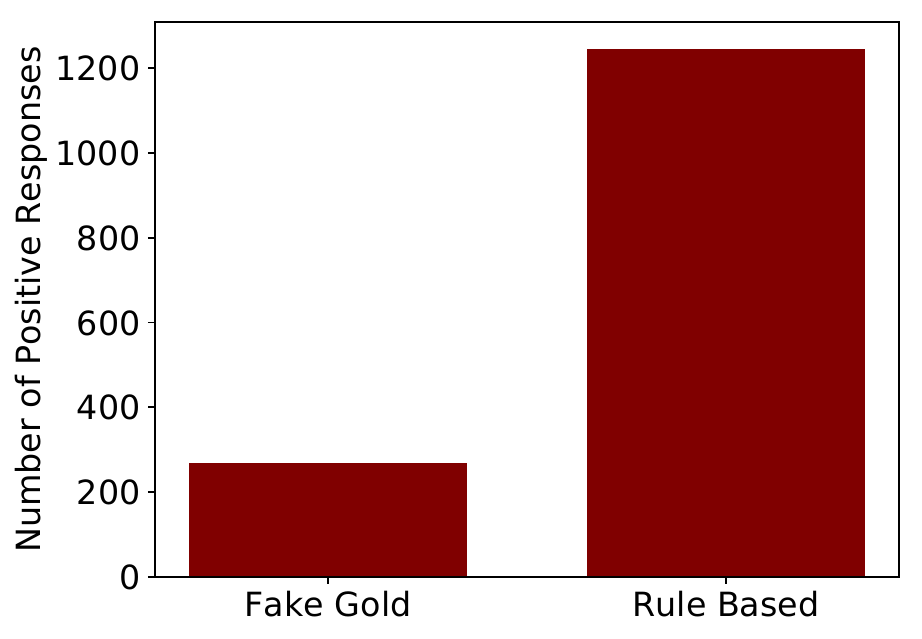}
\caption{Positive responses in our rule-based experiment.}
\label{fig:posrule}
\endminipage\hspace{0.1cm}
\end{figure*}

\textbf{Experiment 1: Tutorials and Gold}
Precision results from our first experiment with 111 HITs and 68 participants, evaluating the impact of different methods of tutorials and gold, are shown in Figure \ref{fig:pre1}.  The number of filtered positive samples were as follows: In the baseline condition we had 37 positive results from 5 workers, in the tutorial overload condition we had 28 positive results from 6 workers, in the gold overload condition we had 6 positive results from 3 workers, and in the Fake Gold condition we had 9 positive results from 6 workers. 

The first thing we can notice is that the precision in the Baseline condition is extremely low (only 27\%).  Clearly, one would not expect such a low base precision for a standard binary labeling task such as sentiment analysis.  This surprising result further motivates our study of task design for constraint tasks, as these tasks are inherently harder for workers due to a variety of factors.

Even more surprising (in light of our hypothesis \textbf{H.1}) is the fact that tutorial overload, which gave workers almost three times the training of the baseline condition, did not improve at all in terms of precision. 
One hypothesis is that upon seeing a large number of questions with given answers, workers begin to think that the task is mostly meant for training, and therefore feel little need to expend much effort.  Alternatively, it may be that these constraint tasks are inherently untrainable, that is, there is some subset of the Turk population who is (for any number of reasons: motivation, time constraints, distractions, etc.) unable to be trained to complete these complex tasks.  

Our data confirms \textbf{H.1} as it relates to the Fake Gold condition. Fake Gold has significantly better precision than both Baseline (p=0.047; FET) and Tutorial Overload (p=0.042; FET). Note that the precision is a substantial jump, from 27\% in Baseline (and 25\% in Tutorial Overload) to 61\% in Fake Gold. Although Fake Gold trended higher than Gold Overload in terms of precision, we did not find any significant difference  (p=0.62; FET).  However, all else being equal, we would much rather choose Fake Gold, as it requires much less expert effort, while allowing workers to do more useful work.



\textbf{Experiment 2: Promoting Careful Thinking}
In the first stage, we found the fake gold condition to be the most promising, so we build conditions on top of Fake Gold (instead of Baseline) in our next experiment.

The precision results from our second experiment\footnote{The experiment was run across two temporally distinct stages, results from the second stage were combined additively with the first stage.} with 263 HITs and 127 participants are shown in Figure \ref{fig:pre2}. ``FG + Explanation'' refers to the one-sided explanation condition in the figure. The number of filtered positive samples were as follows: In the Fake Gold condition we had 23 filtered positive results from 14 workers, in the one-sided explanation  condition we had 15 positive results from 9 workers, in the continuity condition we had 26 positive results from 15 workers, and in the skip button condition we had 33 positive results from 13 workers. 

They show that, in line with our hypothesis \textbf{H.2}, all three methods appear to improve precision over fake gold.

The main result we see in Figure \ref{fig:pre2} is that one-sided explanation seems by far the best, with 87\% precision compared to Fake Gold's 30\%.  This difference is statistically significant (p$<$0.01, FET), demonstrating the benefit of asking the workers to explain the meaning of their ``yes'' answers.

\textbf{Experiment 3: Comparing Explanation Methods}
Our new one-sided explanation method seemed quite promising in experiment 2, but we wanted to make sure that it was more effective than the standard (two-sided explanation) method examined in past work \cite{drapeau2016microtalk}. Therefore, we did a straightforward comparison of the two conditions.

Our third experiment (no figure) had 152 HITs and 92 participants.
The number of filtered positive samples were: 11 positive results from 8 workers in the one-sided explanation condition and 37 positive results from 17 workers in two-sided explanation condition. 

We found that one-sided explanation seemed best, with 73\% precision compared to two-sided explanation which had 45\% precision.  Using a one-tailed-test\footnote{As the point of this experiment was not to test if the two were different, but if one-sided was better than two-sided explanations.} to compare precision, it did not quite rise to the level of statistical significance (p = 0.12, FET).  However, we see a highly significant difference in terms of the percentage of positive responses (p$<$0.001, FET), with one-sided explanation having 10.3\% positive responses (after worker filtering) compared to two-sided explanation which had 32.5\% .  This indicates that workers in the one-sided explanation condition are much more careful about where they select ``yes'' answers, which is desirable behavior in a safety-critical task like this. Also it's important to note that fake gold + one-sided explanation is the only condition to achieve over 65\% precision in any experiment, and has done so in two completely independent experiments.

\textbf{Experiment 4: Rule-Based}
We turn our attention to comparing our rule-based interface to the case-by-case interface.   
The rule-based task took longer for workers to complete than the case-by-case task (and therefore required more pay). This rendered the previous A/B Testing approach infeasible, as Amazon Mechanical Turk does not support programmatically determining how much to pay workers.\footnote{Except through the use of promised bonuses, which did not appear effective in our pilot experiment.}   Therefore, we created separate HIT groups for the two experiments. The HIT groups were released simultaneously, but we consider only results from a single HIT group for each worker. To allow us to better compare the efficiency of the tasks, we decided to equalize the amount we spent on each HIT group: We launched 30 rule-based HITs at a total cost of \$108.6 (prior to Amazon fees), and therefore launched 117 case-by-case HITs for a total cost of \$108.81. 

Additionally, in our Mechanical Turk pilot study we observed substantially decreased results for all interfaces, possibly due to the recently reported rise in bots~\cite{wiredbots}.  Therefore, we increased our task requirements to require 1000 HITs completed and the workers be from the US, Canada, or Singapore.

All tasks completed in less than 24 hours.  More specifically, the case-by-case task took a little over an hour to complete, while the rule-based task took roughly 22 hours. This may have been in part due to the abundance of HITs available on the case-by-case task, as well as the shorter max time allotted (20 minutes instead of 45 minutes) and more lenient task limiting (5 HITs per worker instead of 3). There was likely also an element of self-selection, where workers who were seeking more straightforward work selected other HITs over the rule-based task. Note that self-selection is not necessarily a negative in this situation: we would prefer that workers who (for whatever reason) do not think they can generate high-quality rules to decline the HIT, rather than have to design complicated mechanisms to filter them out after the fact.  

The precision results from our rule-based experiment, with 147 HITs and 64 participants, are shown in Figure \ref{fig:prerule}. Figure \ref{fig:posrule} shows the number of positive samples after filtering:   The case-by-case fake gold condition had 269 positive results from 26 workers, the rule-based condition had 1246 positive results from 7 workers.  Due to the very large number of positive results in certain conditions, we judged the accuracy of only 200 randomly-chosen results, 102 from rule-based and 98 from case-by-case.

We noticed that despite the fake gold case-by-case experiment being virtually identical to that launched in experiments 1 and 2, the precision is much lower than either of those two experiments at 24\%. We are not sure why the quality of work produced by workers on Mechanical Turk has declined in this fashion, our best guess is that our increased qualifications were not entirely successful at preventing the recent influx in bots~\cite{wiredbots} from completing our task.

In any case, we saw the the rule-based condition seemed quite effective when compared to the case-by-case task, even more so than we hypothesized in \textbf{H.3}. As shown in Figure \ref{fig:prerule}, the rule-based condition has higher precision (41\%) than the case-by-case condition (24\%), which is statistically significant (p=0.02, FET).   This suggests that despite the task and interface being significantly more complex, workers were able to write rules that did as good or better than if they were looking at each individual case. We feel this demonstrates both the (often overlooked) ability of crowdworkers to perform complex work, and the successful design of our tutorial and interface in giving workers the appropriate feedback necessary to complete the task. Additionally, because of the power of writing rules, Turk workers were able to label an order of magnitude more states for the same pay, as shown in Figure \ref{fig:posrule}.  

Typically, constraint specification is not a time-critical task, as workers can gradually add constraints to a system over time, and so the extra time taken to get results from the rule-based task is not major concern.  However, our results suggest that our interfaces are perhaps best used in combination: The case-by-case interface can engage a wider group of workers and returns results more quickly, but the rule-based interface is more cost-effective and allows workers to be more efficient with their time while generating high-quality work.


\section{Discussion and Conclusion}
In this paper, we presented one of the first explorations of an important component of AI safety: how to design user interfaces that allow humans to efficiently specify high-quality constraints for real-world AI systems.  
Our results show that, despite the fact that baseline precision is quite low and more training has little impact, our new fake gold and one-sided explanation designs were able to substantially increase precision.  Further, our novel rule-based interface is quite effective, generating better precision than a case-by-case approach while producing an order of magnitude more useful responses.

\begin{figure}
    \centering
    \includegraphics[width=0.5\columnwidth]{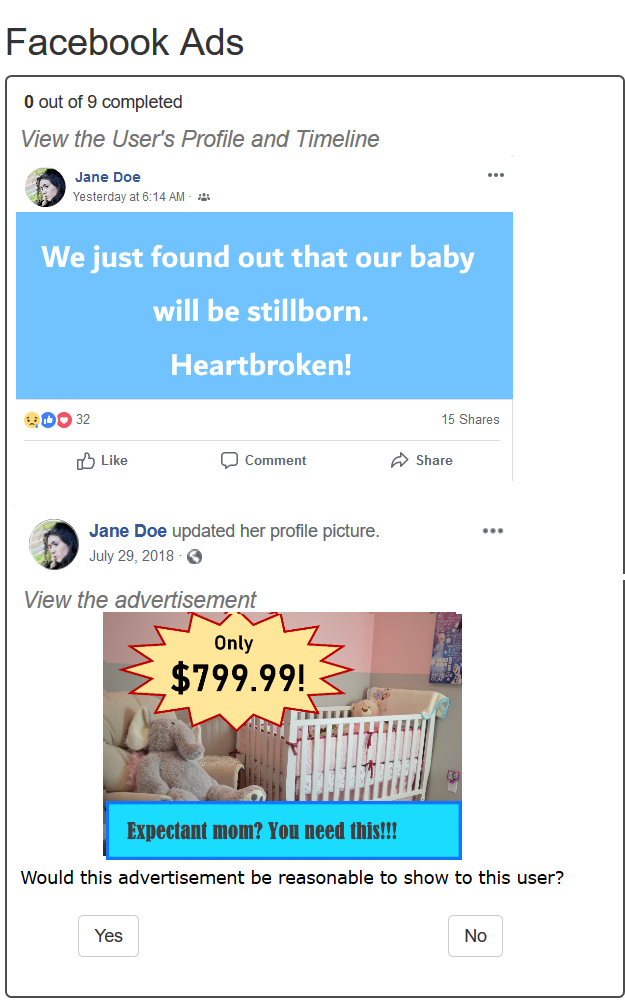}
    \caption{A mockup of our case-by-case interface applied to constraining Facebook's advertisement AI.}
    \label{fig:cbc_fb}
\end{figure}

\begin{figure}
    \centering
    \includegraphics[width=\columnwidth]{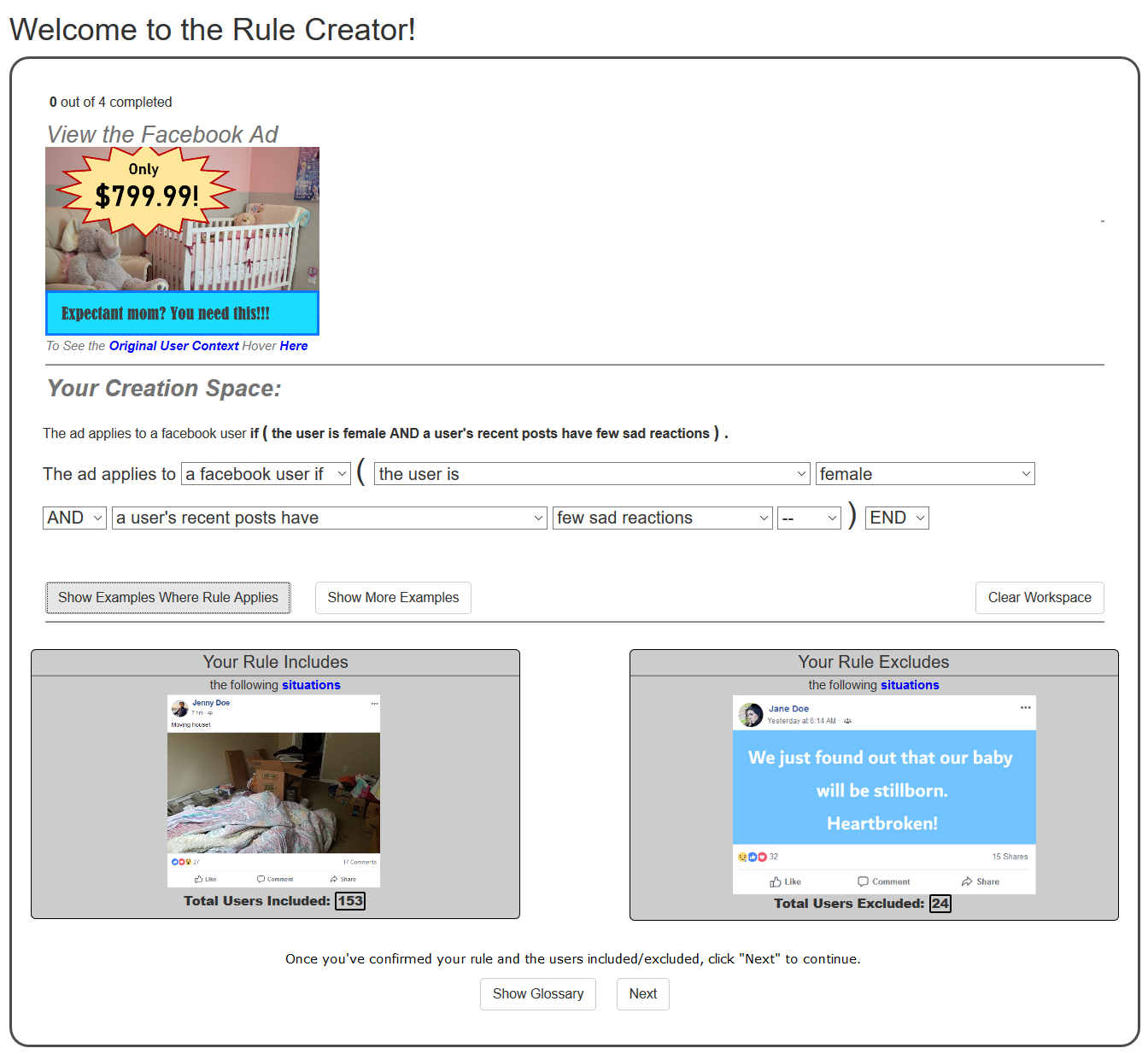}
    \caption{A mockup of our rule-based interface applied to constraining Facebook's advertisement AI.}
    \label{fig:rule_fb}
\end{figure}

Both of our user interface designs are highly general and could be applied to a wide variety of other difficult problems in AI.  For example, there have been recent viral news articles about how Facebook's AI algorithm served a deluge of highly insensitive ads to a mother whose baby was stillborn~\cite{holohanbaby}.  She wondered why Facebook did not use obvious indicators (like the number of sad reactions to her announcement of the loss) to prevent those ads from being shown to her. Our interfaces could easily apply in this setting: states would represent a user and their timeline content, and actions would be potential advertisements.  Figure~\ref{fig:cbc_fb} shows a mockup of what the case-by-case interface might look like in this case, and Figure~\ref{fig:rule_fb} shows a mockup of the rule-based interface. 


Note that, since our maximum precision  was 73\%-86\%, these constraints are not quite ready for direct deployment at the current time.   Future work includes increasing this precision even further through methods like weighted majority voting. 
Despite this, we think these results can already be useful, as experts simply need to confirm the safety of crowdsourced constraints rather than generating brand-new constraints. Another promising direction for future work is exploring methods for combining the rule-based and case-by-case interface to handle challenging state spaces, for example by using the case-by-case interface to generate automatic suggestions (with explanations) on how to improve the rules generated by the rule-based interface.

Although there is much work to be done, in this paper we have taken the first steps towards a future where everyone, regardless of expertise, can be involved in the process of ensuring that AI systems will have the agency to help humankind, but not the agency to harm us.

\section{Appendix}
\subsection{Proof of Theorem 1}
\setcounter{theorem}{0}
\begin{theorem}
Aside from the length limit\footnote{We limit the length of the predicates due to practical storage and data transmission issues.}, our method of rule construction is fully expressive, that is, it can represent any Boolean function over the set of base predicates.
\end{theorem}
\vspace{0.2cm}
\begin{proof}
We will show that we can represent any valid Boolean function in disjunctive normal form (DNF). Recall that in first-order logic, each literal in the DNF is composed of a predicate applied to valid arguments.

First, note that our interface directly allows the user to select predicates and their valid arguments.  Therefore it is possible to construct all valid literals in our interface.

Now, in the DNF, each literal may be negated by the use of the not operator immediately preceding the literal.  Although we do not allow the user to insert an explicit not operator, when choosing predicates we always allow the user to choose its negated form.  Therefore, the user may construct any literal or its negation.

We proceed to show that the user can create any sequence of DNF clauses joined by ORs, by \textbf{induction on the number of literals}.  As part of our inductive hypothesis we also prove an invariant: there is always either exactly one rightmost parenthesis belonging to the current clause\footnote{When we say a parenthesis belongs to the current clause, we mean that it (and it matching parenthesis) does not encompass any literals outside of the clause.} after the last literal $D$, or there is zero and $D$ is the only literal in its clause. 

Note that in a DNF, the result is the same regardless of the order the ORs are evaluated in, due to the well-known associativity of the OR operator. Therefore it is not necessary to show that the clauses are evaluated left-to-right, just that they are fully evaluated (e.g. by being enclosed in parentheses) and then combined with another clause using the OR operator.

Base Case: The dropdowns allow the user to directly create any single literal.  There are no parentheses, but the literal is the only clause member so the inductive hypothesis holds.

Inductive case: Our inductive hypothesis holds for one or more literals and we wish to add an additional literal.  

If there is one literal, our interface treats this differently, AND or OR may be added directly to achieve the desired single-clause or two-clause result. In the case of AND, there is a single rightmost parenthesis and multiple elements in the clause, so the invariant holds.  In the case of OR, there are no right parentheses belonging to the cause, but the last literal is the lone element of the clause so the invariant holds. 

Now, assume there is more than one literal. In that case, there must be at least one logical.  Take the rightmost logical.  By our method of adding parentheses, it must have added a parenthesis just to the right of the last literal $D$. Further, in order to let the user add the next logical (and, subsequently, literal), our method of construction will add a choicebox around this right parenthesis . We can  visually represent this as ``{\dots}D~--)~--\dots'.

Note that, by our inductive hypothesis, either there is exactly one right parenthesis  after $D$  belonging to $D$'s clause or there is zero.  If there is one, it clearly must be the parenthesis immediately following $D$.\footnote{In order for the parenthesis to the right of the parenthesis after $D$ to belong to $D$'s clause, the inner parenthesis just after $D$ would have to as well.}
Therefore,  there are  three cases:
\begin{enumerate}
\item The parenthesis just after $D$ does not belong to the clause, and we want to add the current literal F WLOG, into the same clause as the preceding variable D.  In this case we know D is the sole element of the clause by the inductive hypothesis.   Therefore, we select the inner choicebox and fill it with an AND. Since we chose inner, the left hand parenthesis will go just before D, resulting in ``\dots(D~AND~F))\dots''. Since by our inductive hypothesis D was correctly in a clause by itself before adding $F$, clearly $D$ and $F$ are now correctly in a clause together, as desired.  We have multiple members in a clause and exactly one right parenthesis belonging to the clause to the right of $F$, so the invariant holds.
\item The parenthesis just after $D$ belongs to the clause, and we want to add the current literal F WLOG, into the same clause as the preceding variable D.  In this case one selects the outer choicebox, resulting in ``{\dots}D)~AND~F)\dots''.
Since the  parenthesis just after $D$ belongs to the clause, the matching left paren must be contained within the clause.  The left hand parenthesis will go just before D's left parenthesis, therefore since D was correctly a part of the clause by the inductive hypothesis, F must be as well. We have multiple members in a clause and exactly one right parenthesis belonging to the clause to the right of $F$, so the invariant holds.
\item We want F to be the first element of a new clause.  We select the outer choicebox and fill it with an OR, giving us ``{\dots}D~)~OR~F)\dots''.  Now, as previously shown,   because of the induction hypothesis the right parenthesis immediately following D is the only right parenthesis after D that can possibly belong to the clause.  Therefore, OR and F must correctly be outside of $D$'s clause.  Further, by the inductive hypothesis the DNF expression was built correctly thus far, so if the OR and F are outside of $D$'s clause they must also be outside all clauses.\footnote{Or they would be in a clause that somehow incorrectly has D's clause nested inside of it, violating the inductive hypothesis.}   The only literal in the clause is F, and there are no right parentheses belonging to the clause, so the invariant holds.
\end{enumerate}
\end{proof}

\section{Acknowledgements}
We acknowledge Yvonne Chen, Zoran Popovi\'{c}, Max Panoff, Nick Grogg, Katharina Reinecke, and NSF's RII Track-1: {\okina}Ike Wai: Securing Hawaii's Water Future Award \#OIA-1557349.

\bibliographystyle{aaai}
\bibliography{constraints}

\begin{thebibliography}{}

\bibitem[\protect\citeauthoryear{Ambati, Vogel, and
  Carbonell}{2012}]{ambati2012}
Ambati, V.; Vogel, S.; and Carbonell, J.
\newblock 2012.
\newblock {Collaborative Workflow for Crowdsourcing Translation}.
\newblock In {\em Proceedings of the ACM 2012 Conference on Computer Supported
  Cooperative Work}, CSCW '12,  1191--1194.
\newblock New York, NY, USA: ACM.

\bibitem[\protect\citeauthoryear{Awad \bgroup et al\mbox.\egroup
  }{2018}]{awad2018moral}
Awad, E.; Dsouza, S.; Kim, R.; Schulz, J.; Henrich, J.; Shariff, A.; Bonnefon,
  J.-F.; and Rahwan, I.
\newblock 2018.
\newblock The moral machine experiment.
\newblock {\em Nature} 563(7729):59.

\bibitem[\protect\citeauthoryear{Balakrishnan \bgroup et al\mbox.\egroup
  }{2018a}]{balakrishan2018}
Balakrishnan, A.; Bouneffouf, D.; Mattei, N.; and Rossi, F.
\newblock 2018a.
\newblock Incorporating behavioral constraints in online {AI} systems.
\newblock {\em CoRR} abs/1809.05720.

\bibitem[\protect\citeauthoryear{Balakrishnan \bgroup et al\mbox.\egroup
  }{2018b}]{Balakrishnan2018nflix}
Balakrishnan, A.; Bouneffouf, D.; Mattei, N.; and Rossi, F.
\newblock 2018b.
\newblock Using contextual bandits with behavioral constraints for constrained
  online movie recommendation.
\newblock In {\em Proceedings of the Twenty-Seventh International Joint
  Conference on Artificial Intelligence, {IJCAI} 2018, July 13-19, 2018,
  Stockholm, Sweden.},  5802--5804.

\bibitem[\protect\citeauthoryear{Bernstein \bgroup et al\mbox.\egroup
  }{2015}]{soylent}
Bernstein, M.~S.; Little, G.; Miller, R.~C.; Hartmann, B.; Ackerman, M.~S.;
  Karger, D.~R.; Crowell, D.; and Panovich, K.
\newblock {2015}.
\newblock {Soylent: A Word Processor with a Crowd Inside}.
\newblock {\em {Commun. ACM}} {58}({8}):{85--94}.

\bibitem[\protect\citeauthoryear{Bragg, Mausam, and
  Weld}{2016}]{bragg2016optimal}
Bragg, J.; Mausam, M.; and Weld, D.~S.
\newblock 2016.
\newblock Optimal testing for crowd workers.
\newblock In {\em Proceedings of the 2016 International Conference on
  Autonomous Agents \& Multiagent Systems},  966--974.
\newblock International Foundation for Autonomous Agents and Multiagent
  Systems.

\bibitem[\protect\citeauthoryear{Chen \bgroup et al\mbox.\egroup
  }{2016}]{chen2016crowdsourcing}
Chen, Y.; Mandel, T.; Liu, Y.-E.; and Popovi\'{c}, Z.
\newblock 2016.
\newblock Crowdsourcing accurate and creative word problems and hints.
\newblock {\em AAAI HCOMP}.

\bibitem[\protect\citeauthoryear{Chow \bgroup et al\mbox.\egroup
  }{2018}]{chow2018lyapunov}
Chow, Y.; Nachum, O.; Duenez-Guzman, E.; and Ghavamzadeh, M.
\newblock 2018.
\newblock A {Lyapunov}-based approach to safe reinforcement learning.
\newblock {\em arXiv preprint arXiv:1805.07708}.

\bibitem[\protect\citeauthoryear{Dai, Mausam, and
  Weld}{2011}]{dai2011artificial}
Dai, P.; Mausam, M.; and Weld, D.~S.
\newblock 2011.
\newblock Artificial intelligence for artificial artificial intelligence.
\newblock In {\em Proceedings of the Twenty-Fifth AAAI Conference on Artificial
  Intelligence},  1153--1159.
\newblock AAAI Press.

\bibitem[\protect\citeauthoryear{Dalal \bgroup et al\mbox.\egroup
  }{2018}]{dalal2018safe}
Dalal, G.; Dvijotham, K.; Vecerik, M.; Hester, T.; Paduraru, C.; and Tassa, Y.
\newblock 2018.
\newblock Safe exploration in continuous action spaces.
\newblock {\em arXiv preprint arXiv:1801.08757}.

\bibitem[\protect\citeauthoryear{Davey and
  Priestley}{1990}]{davey1990introduction}
Davey, B.~A., and Priestley, H.
\newblock 1990.
\newblock {\em Introduction to Lattices and Order}.
\newblock Cambridge University Press.

\bibitem[\protect\citeauthoryear{Di~Stefano \bgroup et al\mbox.\egroup
  }{2014}]{stefano2015}
Di~Stefano, G.; Gino, F.; Pisano, G.; and Staats, B.
\newblock {2014}.
\newblock {Learning by Thinking: Overcoming Bias for Action through
  Reflection}.
\newblock {\em {Harvard Business School Working Paper Series}} {58}({14-093}).

\bibitem[\protect\citeauthoryear{Dosovitskiy \bgroup et al\mbox.\egroup
  }{2017}]{dosovitskiy2017carla}
Dosovitskiy, A.; Ros, G.; Codevilla, F.; Lopez, A.; and Koltun, V.
\newblock 2017.
\newblock {CARLA}: An open urban driving simulator.
\newblock {\em arXiv preprint arXiv:1711.03938}.

\bibitem[\protect\citeauthoryear{Drapeau \bgroup et al\mbox.\egroup
  }{2016}]{drapeau2016microtalk}
Drapeau, R.; Chilton, L.~B.; Bragg, J.; and Weld, D.~S.
\newblock 2016.
\newblock Microtalk: Using argumentation to improve crowdsourcing accuracy.
\newblock In {\em Fourth AAAI Conference on Human Computation and
  Crowdsourcing}.

\bibitem[\protect\citeauthoryear{Dreyfuss}{2018}]{wiredbots}
Dreyfuss, E.
\newblock 2018.
\newblock A bot panic hits amazon's mechanical turk.
\newblock Wired Magazine.

\bibitem[\protect\citeauthoryear{Gao \bgroup et al\mbox.\egroup
  }{2015}]{gao2015acquiring}
Gao, J.; Zhuo, H.~H.; Kambhampati, S.; and Li, L.
\newblock 2015.
\newblock Acquiring planning knowledge via crowdsourcing.
\newblock In {\em Third AAAI Conference on Human Computation and
  Crowdsourcing}.

\bibitem[\protect\citeauthoryear{Heer and Bostock}{2010}]{heer2010}
Heer, J., and Bostock, M.
\newblock {2010}.
\newblock {Crowdsourcing Graphical Perception: Using Mechanical Turk to Assess
  Visualization Design}.
\newblock In {\em {Proceedings of the SIGCHI Conference on Human Factors in
  Computing Systems}}, {CHI '10},  {203--212}.
\newblock {New York, NY, USA}: {ACM}.

\bibitem[\protect\citeauthoryear{Holohan}{2018}]{holohanbaby}
Holohan, M.
\newblock 2018.
\newblock Her baby was stillborn, but the ads just kept coming: One mother
  shares her pain.
\newblock Today,
  https://www.today.com/parents/gillian-brockell-s-open-letter-tech-companies-goes-viral-t145124.

\bibitem[\protect\citeauthoryear{Lasecki \bgroup et al\mbox.\egroup
  }{2014}]{lasecki2014using}
Lasecki, W.~S.; Marcus, A.; Rzeszotarski, J.~M.; and Bigham, J.~P.
\newblock 2014.
\newblock Using microtask continuity to improve crowdsourcing.
\newblock Technical report.

\bibitem[\protect\citeauthoryear{Mandel \bgroup et al\mbox.\egroup
  }{2017}]{mandel2017where}
Mandel, T.; Liu, Y.-E.; Brunskill, E.; and Popovi\'{c}, Z.
\newblock 2017.
\newblock Where to add actions in human-in-the-loop reinforcement learning.
\newblock {\em AAAI}.

\bibitem[\protect\citeauthoryear{Mitra, Hutto, and
  Gilbert}{2015}]{mitra2015comparing}
Mitra, T.; Hutto, C.~J.; and Gilbert, E.
\newblock 2015.
\newblock Comparing person-and process-centric strategies for obtaining quality
  data on amazon mechanical turk.
\newblock In {\em Proceedings of the 33rd Annual ACM Conference on Human
  Factors in Computing Systems},  1345--1354.
\newblock ACM.

\bibitem[\protect\citeauthoryear{Mitrovic, Martin, and
  Suraweera}{2007}]{mitrovic2007intelligent}
Mitrovic, A.; Martin, B.; and Suraweera, P.
\newblock 2007.
\newblock Intelligent tutors for all: The constraint-based approach.
\newblock {\em IEEE Intelligent Systems} (4):38--45.

\bibitem[\protect\citeauthoryear{Oleson \bgroup et al\mbox.\egroup
  }{2011}]{oleson2011programmatic}
Oleson, D.; Sorokin, A.; Laughlin, G.~P.; Hester, V.; Le, J.; and Biewald, L.
\newblock 2011.
\newblock Programmatic gold: Targeted and scalable quality assurance in
  crowdsourcing.
\newblock {\em Human computation} 11(11).

\bibitem[\protect\citeauthoryear{O'Sullivan}{2002}]{osullivan2002interactive}
O'Sullivan, B.
\newblock 2002.
\newblock Interactive constraint-aided conceptual design.
\newblock {\em AI EDAM} 16(4):303--328.

\bibitem[\protect\citeauthoryear{Tessler, Mankowitz, and
  Mannor}{2018}]{tessler2018reward}
Tessler, C.; Mankowitz, D.~J.; and Mannor, S.
\newblock 2018.
\newblock Reward constrained policy optimization.
\newblock {\em arXiv preprint arXiv:1805.11074}.

\bibitem[\protect\citeauthoryear{Williams \bgroup et al\mbox.\egroup
  }{2016}]{williams2016axis}
Williams, J.~J.; Kim, J.; Rafferty, A.; Maldonado, S.; Gajos, K.~Z.; Lasecki,
  W.~S.; and Heffernan, N.
\newblock 2016.
\newblock {AXIS}: Generating explanations at scale with learnersourcing and
  machine learning.
\newblock In {\em Learning@ Scale},  379--388.
\newblock ACM.

\bibitem[\protect\citeauthoryear{Zhuo}{2015}]{zhuo2015crowdsourced}
Zhuo, H.~H.
\newblock 2015.
\newblock Crowdsourced action-model acquisition for planning.
\newblock In {\em AAAI},  3439--3446.

\end{thebibliography}

\end{document}